\documentclass[preprint]{revtex4}
\usepackage{graphicx}
\usepackage{dcolumn}
\usepackage{bm}
\usepackage{dsfont}
\usepackage{amsmath}
\usepackage{color}

\begin{document}


\title{Emergence of non-centrosymmetric topological insulating phase in BiTeI under pressure }
\author{M. S. Bahramy$^{1}$}
\email{bahramy@riken.jp}
\author{B. -J. Yang$^{1}$}
\author{R. Arita$^{1,2}$}
\author{N. Nagaosa$^{1,2,3}$}
\affiliation{
{$^1$Correlated Electron Research Group (CERG), RIKEN-ASI, Wako, Saitama 351-0198, Japan}\\
$^2$Department of Applied Physics, University of Tokyo,  Tokyo 113-8656, Japan\\
$^3$Cross-Correlated Materials Research Group (CMRG), RIKEN-ASI, Wako, Saitama 351-0198, Japan\\}
\maketitle
{\bf
The spin-orbit interaction affects the electronic structure of solids in various ways. Topological insulators are one  example where the spin-orbit interaction leads  the bulk bands to have a non-trivial topology, observable as gapless surface or edge states. Another example is the Rashba effect, which lifts the electron-spin degeneracy as a consequence of spin-orbit interaction under broken inversion symmetry.  It is of particular importance to know how these two effects, i.e. the non-trivial topology of electronic states and Rashba spin splitting, interplay with each other. Here we show, through sophisticated  first-principles calculations, that BiTeI, a giant bulk Rashba semiconductor, turns into a topological insulator under a reasonable pressure. This material is shown to exhibit several unique features such as, a highly pressure-tunable giant Rashba spin splitting,  an unusual pressure-induced quantum phase transition, and more importantly  the formation of strikingly different Dirac surface states at opposite sides of the material.}

{\bf Introduction}\\
Theoretical work of Kane and Mele~\cite{kane2005} has marked a milestone in our understanding of the insulating phase of the matter. In their seminal work, they  classified the bulk insulators according to the topological order of their electronic states using quantum metrics known as topological indices. When nontrivially ordered, it was predicted that  such  topological insulators (TI's) should exhibit gapless states at their boundaries, e.g. at the surface or edges.  The succeeding works have confirmed this prediction by theoretically proposing and/or  experimentally discovering a number of two and three dimensional TI's, including  HgTe quantum wells~\cite{konig2007,bernevig2006} and uniaxially strained bulk HgTe~\cite{fu2007,brune2011}, a number of layered V$_2$VI$_3$ binary compounds e.g. Bi$_2$Se$_3$, Bi$_2$Te$_3$, Sb$_2$Te$_3$~\cite{zhang2009,xia2009}, and  the ternary compounds LaBiTe$_3$~\cite{yan2010}, TlBiTe$_2$, and TlBiSe$_2$~\cite{chen2010,liu2010}. The recent first-principles studies have further  proposed  that several half-Heusler compounds such as LaPtBi, LuPtSb, and YPdBi~\cite{xiao2010,feng2010,chadov2010,lin2010} can  turn into a TI under a uniaxial strain, similar to what was predicted and observed for HgTe~\cite{fu2007,brune2011}.  Common to all these materials is the presence of heavy elements with a reasonably large atomic spin-orbit interaction (SOI). Such a similarity arises from the fact that in all TI systems the bulk band gap is the result of band inversion by SOI~\cite{fu2007}. In other words, without such a SOI-induced band inversion, no topological phase can be realized in a system.

 In a more general context, SOI tends to lift the degeneracy of energy bands. In systems with inversion ($I$) symmetry, such a splitting leaves the energy states spin-degenerate, as long as the time-reversal ($T$) symmetry is hold. $T$-symmetry connects
the states $\psi_{\bf{k}, \uparrow}$ and $\psi_{-\bf{k}, \downarrow}$, while the $I$-symmetry  enforces the degeneracy between $\psi_{\bf{k}, \uparrow(\downarrow)}$ and $\psi_{-\bf{k}, \uparrow(\downarrow)}$. Breaking $I$ symmetry lifts the latter constraint and, hence, lets the energy bands  be spin-split at generic $k$-points.
Rashba spin splitting (RSS)~\cite{rashba} well exemplifies this situation, as described by $H_R = {{ \bf{p}^2} \over {2 m}} + \nu\lbrack \bf{e} \cdot ( \bf{s} \times \bf{p})\rbrack$
where $\bf{e}$ is the direction of the potential gradient, breaking $I$-symmetry,
and $\bf{s}$ and $\bf{p}$ are the spin and momentum operators, respectively.
This interaction leads to several unique phenomena, such as the
spin Hall effect~\cite{sinova2004}, the spin Galvanic effect~\cite{ganichev}, and the magneto-electric effect~\cite{chalaev}.
Furthermore, once  superconductivity occurs in a Rashba system,
unusual features such as the mixing of the singlet and triplet pair states,
a large upper critical field beyond the Pauli limit~\cite{bauer,frigeri}, and
topological superconductivity  with Majorana edge channels~\cite{tanaka} can
appear.  As SOI is responsible for both RSS and non-trivial topology of electronic states in semiconductors, it is therefore of fundamental  interest to know how they interplay with each other once they coexist in a  system and more importantly what the impact of such an interplay is on the electronic structure of bulk and boundary states. 

The purpose of this work is to  study such an interplay in a layered  polar compound, BiTeI. Backed by our earlier band structure calculations~\cite{ishizaka2011,bahramy2011}, the angle resolved photoemission spectroscopy (ARPES) measurements~\cite{ishizaka2011,wray} have  revealed that the bulk  conduction and valence states in BiTeI are subject to a giant RSS of the order of several hundred meV, lying among the highest reported so far. In this work we further show using  the first-principles calculations that by applying a reasonable hydrostatic pressure, $P$, the material turns into a TI with many interesting features, such as a nearly double enhancement in RSS accompanied with an unconventional metallic behavior at quantum phase transition. It is also demonstrated that unlike the $I$-symmetric TI's, the gapless  Dirac states in TI phase of BiTeI  have completely different shapes at different sides of the material. Consequently, here the surface states on both sides can interestingly have  the same spin helicity in a rather broad range of energies inside the bulk band gap, in sharp contrast with the centrosymmetric TI's.

{\bf Results}\\
{\bf Basic properties.} BiTeI belongs to the trigonal space group of $P3m1$. As shown in Figs. 1-(a) and 1-(b), the crystal structure of BiTeI has a non-centrosymmetric  layered structure along its crystallographic $c$-axis with three atoms in one unit cell. Within each unit, a Bi atom is sandwiched between one Te and one I, forming a triple layer.  Due to the strong covalency and ionicity of Bi-Te and Bi-I bonds, respectively, the bulk crystal intrinsically possesses a  polar axis along the $z$-direction. Despite the strong chemical bonding within each triple layer, the adjacent triple layers are weakly coupled via van-der Waals interaction. From our previous calculations~\cite{bahramy2011}, we know that around the Fermi level $E_F$, all the bands are essentially $p$-type, with conduction bands dominated by Bi-$6p$ and  valence bands composed of Te-$5p$ and I-$5p$ states.  Moreover, due to the negative  crystal field splitting (CFS) of the  valence  bands and positive CFS of conduction bands near $E_F$, in the absence of SOI the top valence bands (TVB's) and bottom conduction bands (BCB's) both become $p_z$ type (to avoid any confusion, hereafter they are referred to as $p_z^A$ and $p_z^B$, respectively). These  features make the electronic structure of BiTeI very much similar to that of the well known TI systems Bi$_2$Te$_3$ and  Bi$_2$Se$_3$~\cite{zhang2009}. However, here due to the absence of $I$-symmetry, it's not possible to assign a distinct even or odd parity to each band.  It is also worth noting that in the case of BiTeI, the minimum energy gap $E_G$ is  not at the Brillouin zone (BZ) center $\Gamma$ but areound the hexagonal face center of BZ, $A$ point,  where $k_x=k_y=0$ and $k_z=\pi/c$, as shown in Fig. 1-(c). As will be shown below, because of the latter difference, BiTeI in TI phase shows a rather different topological order from what has been found for Bi$_2$Te$_3$ and  Bi$_2$Se$_3$.  Introducing SOI, both spin and orbital mixings are allowed. Consequently, $p_z^A$ ($p_z^B$) transforms to $\arrowvert p^A,\pm\frac{1}{2}\rangle$ ( $\arrowvert p^B,\pm\frac{1}{2}\rangle$), thereby getting  energetically repelled upward (downward). This accordingly closes the band gap from 1.2 eV down to $0.286$ eV and induces a giant bulk RSS  among these two sets of bands around point $A$ (see discussion in Ref.~\cite{bahramy2011}). Despite such a huge reduction in $E_G$, BiTeI remains a trivial insulator as $E_G$ still originates from atomic orbital hybridization between Bi and its neighboring Te and  I atoms. The respective band diagram is shown in Fig. 1-(d).

{\bf Topological phase transition in BiTeI under pressure.} Our strategy to turn BiTeI into a TI is to modify its chemical bonds by applying an external hydrostatic pressure. As schematically shown in ~Figs. 1-(d), 1-(e) and 1-(f), through this modification we can effectively control both CFS and SOI such that at a critical pressure $P_c$,  $\arrowvert p^A,\pm\frac{1}{2}\rangle$ and $\arrowvert p^B,\pm\frac{1}{2}\rangle$ become degenerate, whereas at higher $P$'s their energy ordering is reversed thereby forming an inverted band gap. Controlling CFS by $P$ is rather easy to understand, because any change in Bi-Te and Bi-I bonds leads to a change in the energy splitting of  $p^{A,B}_z$ and $p^{A,B}_{x,y}$ states. For example, our non-relativistic band structure calculations reveal that at $P_c$, the CFS of TVB's is so enhanced that, $E_G$ is reduced by 200 meV (see the Supplementary  Fig. S1). Any band gap narrowing associated with CFS modification can be further enhanced  through a subsequent enhancement of RSS of $\arrowvert p^A,\pm\frac{1}{2}\rangle$ and  $\arrowvert p^B,\pm\frac{1}{2}\rangle$ states. As described in detail in Ref.~\cite{bahramy2011}, this is due to the fact that these two states are symmetrically of the same character and hence can very effectively couple with each other through a Rashba-type  Hamiltonian if they are energetically close to each other. In other words, the closer they are to each other, the larger RSS would be achieved.

To elucidate this mechanism,  we show in Figs. 2-(a), 2-(b) and 2-(c) the respective electronic band dispersions of TVB's and BCB's along the high symmetry direction $H-A-L$ of BiTeI as hydrostatically compressed by $V/V_0=1$, $V/V_0=0.89$ and $V/V_0=0.86$, where $V_0$ corresponds to the lattice volume at the ambient pressure $P_{ambient}$.  As shown, at $P_{ambient}$ $(V/V_0=1)$ a comparable giant RSS can be seen for both sets of bands with an $E_G$ of 286 meV. The corresponding Rashba energy $E_R$, defined as the energy difference between the conduction band minimum (CBM) and the conduction band crossing point, is nearly 110 meV, in perfect agreement with that observed by ARPES measurements~\cite{ishizaka2011}.
Compressing the volume down to $V/V_0=0.89$, $E_R$  monotonically increases until it reaches $~200$ meV, astonishingly about two times larger than that  at $P_{ambient}$, as shown in  Fig. 2-(b) (for a detailed  comparison see the Supplementary  Fig. S2, also).  At this point, the system reaches its quantum critical point, represented by a full band gap closing along the $A-H $ directions. It is worth noting that such a quantum phase transition in BiTeI differs from that in usual centrosymmetic TI's, as there it is mediated through a band gap closing at a single high symmetry $k$-point, e.g. $\Gamma$ point~\cite{xu,sato}, whereas in BiTeI as will be shown shortly due to the spin splitting the band gap is closed at six $k$-points, all along $A-H$ directions. Thus, they are describable by six $2\times 2$ Dirac Hamiltonians.  Here, it is important to stress that because of $T$-symmetry these six $k$-points are two-by-two paired with each other, therefore their corresponding states form three pairs of $2\times 2$ Dirac fermions. Since the band touching is mediated through {\it odd} number of such pairs, the transition to TI phase is in principle allowed. 
 Further increasing $P$,  $E_R$ starts decreasing and an inverted band gap  emerges between TVB's and BCB's. Figures 2-(d), 2-(e) and 2-(f) demonstrates the orbital characteristics of BCB's and TVB's at $V/V_0=1$, $V/V_0=0.89$ and $V/V_0=0.86$, respectively. A thorough   analysis of  these bands reveals a  clear change in their atomic orbital characteristics  as a function of $V/V_0$. While at $V/V_0= 1$ the BCB's and TVB's are dominated by Bi-$6p$ and Te-$5p$, respectively, at $V/V_0=0.89$ they appear to share rather the same orbital characters near the band touching point. On the other hand at $V/V_0=0.86$, one can clearly see that Bi-$6p$ orbitals now contribute more significantly  to TVB's whereas the BCB's become strongly dominated by Te-$5p$ and I-$5p$ states, a clear indication that a band inversion has happened and, hence, the system is now in the TI phase. It is worth adding that in other parts of BZ including the plane encompassing $\Gamma$, $M$ and $K$ the band gap remains finite for the whole range of $P$'s. 

To make sure that BiTeI for $P>P_c$ becomes a TI, we have calculated the so-called $\mathds{Z}_2$ topological invariant of the whole system when hydrostatically compressed by $V/V_0=0.86$. By  definition, for three-dimensional systems,  $\mathds{Z}_2$ topological invariant is determined by four indices, among which one is called the strong topological index $\nu_0$ and the three others are weak topological indices $\nu_1$, $\nu_2$ and $\nu_3$~\cite{fu2007}. The complete list is then described as $\mathds{Z}_2=\nu_0;(\nu_1\nu_2\nu_3)$.  For a normal insulator all the four indices are zero whereas a strong TI is characterized by $\nu_0=1$. If $\nu_0=0$ but any of $\nu_{1-3}$ is 1, the corresponding system is called a weak TI. For the numerical computation of $\mathds{Z}_2$ invariants, several approaches have been proposed~\cite{soluyanov,yu,ringel,fukui}. Here we follow the approach introduced in Ref~\onlinecite{soluyanov} (see the Supplementary Methods and Supplementary Fig. S3 for a full description of our computational method  for the calculation of  the $\mathds{Z}_2$ topological invariants).  Our calculations show that at $V/V_0=0.86$, BiTeI belongs to $\mathds{Z}_2$=1;(001). This clearly indicates that it is a strong TI. Interestingly, the corresponding $\nu_3$ turns out to be 1, which can be attributed to the band touching in the hexagonal face of BZ with $k_z=\pi/c$, encompassing $A$ and $H$ points.  
The interference between the physics of strong and weak topological insulators could be therefore an interesting problem here. For example, $\nu_3=1$ means that the one-dimensional (1D) conduction channel can appear by dislocation with a Burger's  vector along the $z$-axis~\cite{ying2009}. Thus, the present system may offer an ideal medium to study the interaction between the surface Dirac fermion as guaranteed by $\nu_0=1$ and such a 1D channel.

To further confirm that BiTeI becomes a TI for $P>P_c$, we next show in Fig. 3 the surface band structures corresponding to I- and Te-terminated sides of BiTeI. While  the surface states are fully gaped at $P_{ambient}$,  a gapless state appears for the both sides at $P_c$.   At higher pressures, as shown for the case of $V/V_0=0.86$, gapless surface states appear within the bulk band gap. A clear indication that BiTeI has now become a strong TI and, thereby  further confirming the results of our $\mathds{Z}_2$ invariants calculations.  Interestingly, the shape of the Dirac surface states at the Te-ended side of BiTeI completely differs from that at the I-ended one.  While for the former the Dirac point is deeply buried inside the  energy valley formed by the Rashba-split TVB's (resembling the Dirac surface states in Bi$_2$Te$_3$),  on the other side the Dirac point is energetically  well above the bulk CBM. By increasing $P$, due to the reduction of RSS and the subsequent increase in $E_G$, the latter Dirac point becomes energetically  closer to CBM. We expect  this to be a characteristic feature for all non-centrosymmetric TI candidates making them  distinguishable from $I$-symmetric TI systems. As a consequence, in a non-centrosymmetric TI, 
the electrons are fractionalized into nonequivalent halves on top and bottom surfaces. This would lead to some novel features and new spintronics functionalities. As shown in Fig. 3-(g) and 3-(h), the surface Dirac fermions show the similar spin polarization patterns at the top and bottom surfaces, in sharp contrast to the centrosymmetric TI's. It is to be noted that for some thin slabs of Bi$_2$Se$_3$ and Bi$_2$Te$_3$, such an effect appears to happen when the equivalence of the two surfaces is lifted, e.g. by applying an electric field~\cite{yazyev2010} or by differing chemical functionalization~\cite{jin2011}.   Because of this effect,  the Dirac point traverses across the bulk band gap on going from the top side to the bottom side, and always crosses $E_F$ on the side surface as long as $E_F$ is within the bulk gap. Therefore, if one applies a magnetic field or dopes  magnetic impurities, an insulating  stripe is expected to appear on the side surface  (since $E_F$ is now located inside the energy gap of the side surface Dirac fermion). Also the giant spin Galvanic effect is expected since the current-spin relation of the top side is the same as taht of the bottom side. When the hybridization between the Dirac fermions at top and bottom surfaces occurs, in the momentum space a gap appears along a one-dimensional (nearly) circular path, resulting in the divergence of the  density of states. This accordingly leads to the enhanced electron correlation effect and, consequently, to the excitonic instability. These are just a few examples of the novel phenomena expected in the non-centrosymmetric TI's.

{\bf Estimation of $P_c$.} Let us now briefly discuss on the possible value of $P_c$ required for topological phase transition. Performing two  sets of volume optimization calculations using the local density approximation (LDA) and generalized gradient approximation (GGA) and then fitting the respective free energies $E(V)$ to the Murnaghan equation of state~\cite{murnaghan} 
\begin{equation}
\label{murnaghan}
E(V)=E_0+\frac{B_0V}{B_0^\prime}\left [ \frac{(V_0/V)^{B_0^\prime}}{B_0^\prime-1}+1\right ]-\frac{B_0V_0}{B_0^\prime-1}
\end{equation} 
(see Fig. 4), we estimate the upper and lower limits of Bulk modulus $B_0$ of BiTeI to be $~\sim21.9$ GPa and $~\sim 8.9$ GPa. The respective LDA and GGA values of pressure derivative of bulk modulus $B_0^\prime$ are similarly found to be 7.7 and 7.8. As mentioned above, our calculations indicate that at $P_c$,  $V$ is compressed by $~11$\%. Using the relation~\cite{murnaghan}
\begin{equation}
\label{murnaghan2}
 P(V)=\frac{B_0}{B_0^\prime}\left [ \left(\frac{V_0}{V}\right)^{B^\prime_0}-1\right],
 \end{equation}
$P_c$ is accordingly expected to be in the range of 1.7 GPa to 4.1 GPa. Given that this range of $P$'s is rather easily affordable in laboratory, we thus hope this work would stimulate   experimentalists  in this field to explore such an intriguing topological insulating phase in BiTeI.

{\bf Discussion}\\
At this point it is worth explaining (i)  as to why BiTeI becomes gapless only at a certain $P_c$ but not  for a range of $P$'s as discussed before~\cite{Murakami, Murakami2}  and (ii) why such a gapless state occurs along a specific direction. Group theory is the key to answer these questions. Let us first explain the role of crystalline symmetry for the development of dispersion minima (maxima)
 at BCB (TVB) along particular directions. As already mentioned, BiTeI belongs to $C_{3v}$ symmetry consisting of a three-fold rotation $C_3$ along the $z$-direction and three mirror operations $M$: $y\to -y$ where $y$ is along $A-H$ directions. For spin 1/2 electrons, $C_3$ and $M$ can be represented as $e^{-i\sigma_z\pi/3}$ and $i\sigma_y$, respectively, where $\sigma_{x,y,z}$ are Pauli matrices for spin degrees of freedom. Additionally  $T$ operator can be defined as $i\sigma_yK$, where $K$ is the complex conjugation. We can then construct a two band Hamiltonian $H_c(k)$ for the BCB's invariant under $C_3$, $M$ and $T$. Up to cubic  terms of $k$ it turns out to be: 
 \begin{equation}
\label{hk}
 H_c(k)=\frac{k_x^2+k_y^2}{2m^*_{\perp,c}}+\frac{(k_z-\pi/c)^2}{2m^*_{\parallel,c}}+\nu_{k,c}(k_x\sigma_y-k_y\sigma_x)+\lambda_c(3k_x^2-k_y^2)k_y\sigma_z,
 \end{equation}
where $m^*_{\perp,c}$ and $m^*_{\parallel,c}$ are the  in-plane and out-of-plane effective masses of BCB's and $\nu_{k,c}=\nu_c (1+\alpha_c k^2)$. Note that in a similar manner, one can construct $H_v(k)$ for TVB's. The third term in $H_c(k)$ is obviously the Rashba term allowing a cylindrical  in-plane spin splitting within $k_{x,y}$  plane. The fourth term, hereafter referred to as $H_w(k)$, acts as a warping term, trigonally distorting the energy bands.
 
Due to $H_w(k)$,  the  inner and outer branches of the Rashba-split conduction bands  are  distorted in a way that by approaching CBM they first merge together at $k$-points along $A-L$ directions and then form six energy pockets  each of which centered along one  of six  $A-H$ directions, as shown in  Fig. 5-(a). Exactly at CBM  these six pockets reduce to six points. In the same manner,  $H_w(k)$ distorts TVB's such that the valence band maximum (VBM) appears as a point similarly lying along $A-H$. This accordingly explains why band touching between TVB's and BCB's occurs along $A-H$ direction at $P_c$. It is to be noted that as a direct result of  $H_w(k)$, the energy gap along $A-L$ directions is always larger than that along $A-H$ directions for the whole range of $P$'s, as shown in Fig. (2) and also the Supplementary Fig. S2-(e). Accordingly, neither at $P_c$ nor at any other $P$ the band touching can  happen along $A-L$ directions.   The effect of $P$ is to enhance $\lambda$ for $P\leq P_c$ such that at the band touching points one can clearly see a rather large gap (as large as $~40$ meV) along $A-L$ directions where $H_w$ contribution is zero. Such a warping effect and its enhancement at quantum critical point can be well understood by comparing  Figs. 5-(b), 5(c) and 5-(d), in which the isocontours of energy for an arbitrary $E_F$ fixed at 20 meV above the CBM (corresponding to the dashed brown lines in Fig. 2) are shown at different $P$'s. As can be seen, at $P_c$ the inner and outer branches of Rashba-split conduction bands are just about to form the six energy pockets, whereas away from $P_c$ they form two distinct rings among which the outer one appears to be more significantly distorted by $H_w$.

 Let us now address the first question, that is, why the gapless state in BiTeI can be realized only at a certain $P_c$ but not for a range of $P$'s.  For the description of gap-closing we can focus on the two bands touching at $P_{c}$
among the four bands near the chemical potential. Generally, the topological phase transition in time-reversal invariant non-centrosymmetric systems can be described by using a
two-band Hamiltonian $H_{2\times 2}(\textbf{k},P)=\sum_{i=0}^{3}f_{i}(\textbf{k},P)\tau_{i}$
where $\tau_{1,2,3}$ are the Pauli matrices and $\tau_{0}$ is the unit matrix.
The real functions $f_{0,1,2,3}$ depend on the three momenta $\textbf{k}=(k_{x},k_{y},k_{z})$ and  $P$, and are assumed to take into account the interaction of conduction and valence bands.
The topological phase transition through an accidental band touching
occurs if and only if the three conditions of $f_{1,2,3}(\textbf{k},P)=0$ are satisfied
simultaneously in the ($\textbf{k}$, $P$) space.
According to the recent work by
Murakami~\cite{Murakami, Murakami2},
the band touching points, in general,
form a curve in the ($\textbf{k}$, $P$) space
because the three conditions $f_{1,2,3}(\textbf{k},P)=0$
cannot uniquely specify the four parameters ($\textbf{k}$, $P$).
Therefore if the system is free of additional constraints other than the time-reversal symmetry,
a gapless metallic phase is expected to appear in a finite range of $P\in(P_{c1},P_{c2})$
between the two critical pressures $P_{c1}$ and $P_{c2}$.
The occurrence of the single  $P_{c}$ in BiTeI is traced back to
the fact that the band touching occurs along a particular direction in BZ on which
the Hamiltonian has an additional symmetry constraint.
Along the $A-H$ line, e.g., $(k_{x},k_{z})=(0,\pi/c)$, the system is invariant under the combined operation $\Omega\equiv TM$
of the $T$ and $M$ symmetries.
For $T=i\sigma_yK$ and $M=i\sigma_{y}$,
$\Omega$ is given by $K$, which imposes
the following reality conditions: $\Omega H_{2\times 2}(k_{y},P)\Omega^{-1}=H_{2\times 2}^{*}(k_{y},P)=H_{2\times 2}(k_{y},P)$.
Because of this reality condition, $f_{2}(k_{y},P)=0$
at all points along the $A-H$ direction.
Therefore the band touching can be achieved when the two conditions
$f_{1,3}(k_{y},P)=0$ are satisfied in the $(k_{y},P)$ space.
Since the number of conditions to be satisfied is the same as the number of parameters,
a gapless phase appears only at a single critical point $(k_{y,c},P_{c})$.

To conclude, using the first-principles calculations we have examined the role of spin-orbit interaction in the giant bulk Rashba semiconductor BiTeI and found that under the application of pressure, it leads the system to be an unusual topological insulator. The novel surface states and quantum critical phenomena are intriguing problems for further studies. The effects of the long range Coulomb interaction and disorder are, for example, two important issues remaining to be addressed by both the theory and experiment. 
  \newpage 
 {\bf Methods}\\
 {\bf Volume optimizations.} To simulate the effect of $P$, we  optimized the structure of BiTeI crystal at various volumes ranging from $V/V_0=1$ to $V/V_0=0.84$, where $V_0$ denotes the volume of BiTeI unitcell at  $P_{ambient}$ ($a=4.339$ \AA~ and $c=6.854$ \AA). For a given volume, both the atomic positions and crystal's shape were allowed to be fully optimized until the magnitude of force on all ions became less than 0.005 eV/\AA. All the structural optimization calculations were performed using  both the LDA functional and GGA functional of Perdew-Burke-Ernzerhof (GGA-PBE) as implemented in the VASP program ~\cite{vasp,vasp2}.
 The corresponding Brillouin zone was sampled by a $20\times 20\times 20$ $k$-mesh. The respective upper and lower limits of bulk modulus $B_0$  was estimated by fitting the LDA and GGA-PBE free energies $E(V)$ to the Murnaghan equation of state ~\cite{murnaghan}, using Eqn.~\ref{murnaghan}. To relate the volume changes to $P$, we then used Eqn.~\ref{murnaghan2}. 
 
 {\bf Bulk electronic structure calculations.} Within the same level of GGA-PBE theory, the electronic structures of the optimized structures were calculated using the augmented plane wave plus atomic orbitals (APW-LO) method as implemented in the WIEN2K program ~\cite{wien2k}. 
 For this calculations, the muffin tin radii were set to  $R_{MT}=2.5$ bohr for all the atoms and the maximum modulus of the reciprocal vectors $K_{max}$ was chosen such that $R_{MT}K_{max}=7.0$.

 {\bf Surface electronic structure calculations.} To calculate the surface band structure, for each $V/V_0$ the corresponding APW-LO Hamiltonian was first downfolded into an effective low energy $18\times18$ tight binding model  using maximally localized Wannier functions (MLWF's)~\cite{souza,mostofi,kunes}. We chose $p_x$, $p_y$ and $p_z$ states as the projection centers on  all I, Te and Bi atoms to span the top 12 valence bands and the 6 lowest conduction bands by our $18\times18$ models. Based on these realistic models, we then constructed large tight binding supercell Hamiltonians composed of 50 units of BiTeI along its hexagonal $c$-axis. The surface states of each side is then computed by diagonalizing the whole Hamiltonian and projecting the corresponding eigenstates onto the MLWF's of the corresponding surface layer.

{\bf Acknowledgment}\\
 This research is granted by the Japan Society for the Promotion of Science (JSPS) through the "Funding Program for World-Leading Innovative R\&D on Science and Technology (FIRST Program)", initiated by the council for Science and Technology Policy (CSTP). M.S.B. gratefully acknowledges A. A. Soluyanov for his helpful feedbacks.\\
 


\clearpage

\noindent {\bf Fig. 1:} {\bf Crystal structure, Brillouin zone and schematic diagram of band splitting.} (a) Crystal structure of BiTeI. A triple layer with Te-Bi-I is indicated by the purple square. (b) The relative in-plane positions of Bi, Te and I as seen along the $z$ axis. (c) The corresponding Brillouin zone. The pressure-induced evolution of BiTeI from a trivial insulator to a topological insulator is schematically  drawn in (d-f). In each panel, the evolution of atomic p orbitals into the conduction and valence bands of BiTeI is described as 
the crystal field splitting and spin-orbit coupling are turned on in sequence. As shown in (d) below  $P_c$, the bulk band gap $E_G$ is trivial due to the chemical bonding. At $P_c$, as depicted in (e) a gapless state is realized.  Eventually, for pressures beyond $P_c$, the band diagram in (f) shows a band inversion around $E_F$, characterizing topological insulating phase in BiTeI. Here $\pm 1/2$ and $\pm 3/2$ indicate the $z$-component of the total angular momentum.\\
\\
\noindent {\bf Fig. 2:} {\bf Effect of pressure on the bulk electronic states.} Electronic  dispersions of Rashba-split   BCB's and TVB's in BiTeI as hydrostatically compressed by  (a) $V/V_0=1$ , (b) $V/V_0=0.89$ and (c) $V/V_0=0.86$. As shown at $V/V_0=0.89$ a band touching between BCB's and TVB's occurs along $H-A$ direction.   The dashed (brown) line indicates the position of a chemical potential 20 meV above the corresponding CBM of each case (see the related discussion).  In (a) the spin directions are defined such that they are perpendicular to the $k_z$ axis as well as their corresponding $k$-vector. The orbital characteristics of BCB's and TVB's of (a), (b) and (c) are shown in (d), (e) and (f), respectively. The red, blue and green balls correspond to Bi-$6p$, Te-$5p$ and I-$5p$ states, respectively. \\
\\
\noindent {\bf Fig. 3:} {\bf Effect of pressure on the surface states.} Electronic band dispersions near $E_F$ as obtained for the I-terminated side (top panels) and Te-terminated side (bottom panels) of BiTeI, hydrostatically compressed by   (a-b) $V/V_0=1$ , (c-d) $V/V_0=0.89$ and (e-f) $V/V_0=0.86$. The  Fermi surfaces corresponding to  (e) and (f) are shown in (g) and (h), respectively. As depicted schematically in the insets, for an arbitrary $E_F$ located at the middle of bulk gap, the Fermi surface at I-terminated side has a completely  different shape from that of Te-terminated side, but interestingly for both sides, the same spin helicities are seen.\\
\\
\noindent{\bf Fig. 4:} {\bf Volume-dependence of the free energy.} Volume-dependence of free energy $E(V) $ as obtained from (a) GGA and (b) LDA  calculations. To estimate the upper and lower limits of bulk modulus $B_0$ and the corresponding  pressure derivatives of bulk modulus $B_0^\prime$ , the corresponding GGA and LDA data are fitted to the Murnaghan equation of state (solid line).\\
\\
\noindent{\bf Fig. 5:} {\bf Energy evolution of the conduction bands.} (a) Energy isocontours corresponding to various chemical potentials above CBM of BiTeI with $V/V_0=0.89$.   In (b), (c) and (d), the respective energy spectra of BCB's mapped into the $A$ plane ($k_x$,$k_y$,$k_z=\pi/c$) of BiTeI with $V/V_0=1$,  $V/V_0=0.89$ and $V/V_0=0.86$ are shown.  The upper limit of energy (represented by the red color)  corresponds to the Rashba energy $E_R$ of BCB's at the given $V/V_0$ . The dashed lines indicate the isocontours of energy for a chemical potential fixed at 20 meV above the CBM of each  case. Arrows denote the spin directions.  \\

\newpage

\begin{figure}[pt]
\begin{center}
\rotatebox{0}{\includegraphics[width=5 in]{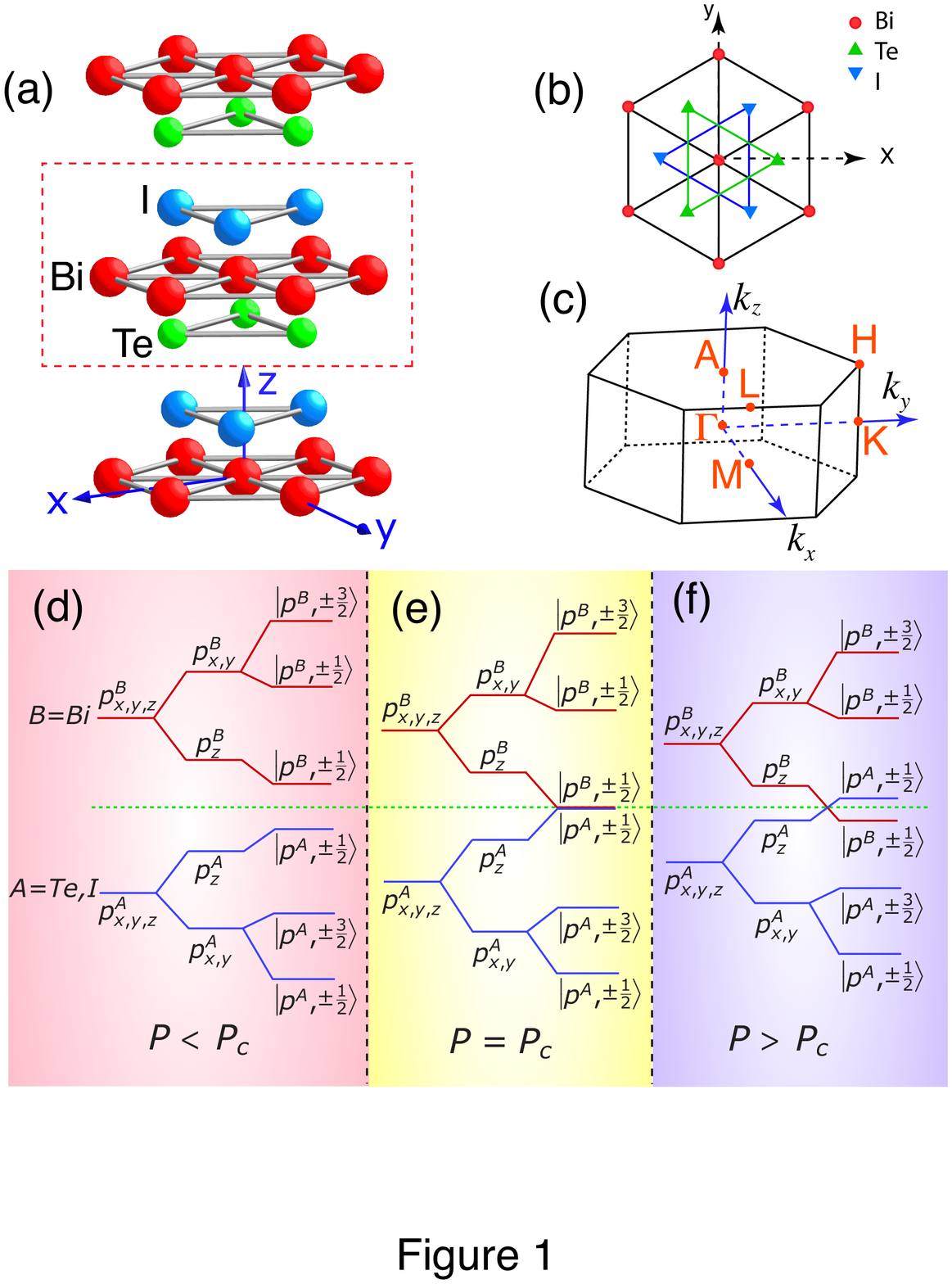}}
\end{center}
\end{figure}
 
\newpage

\begin{figure}[pt]
\begin{center}
\rotatebox{0}{\includegraphics[width=5 in]{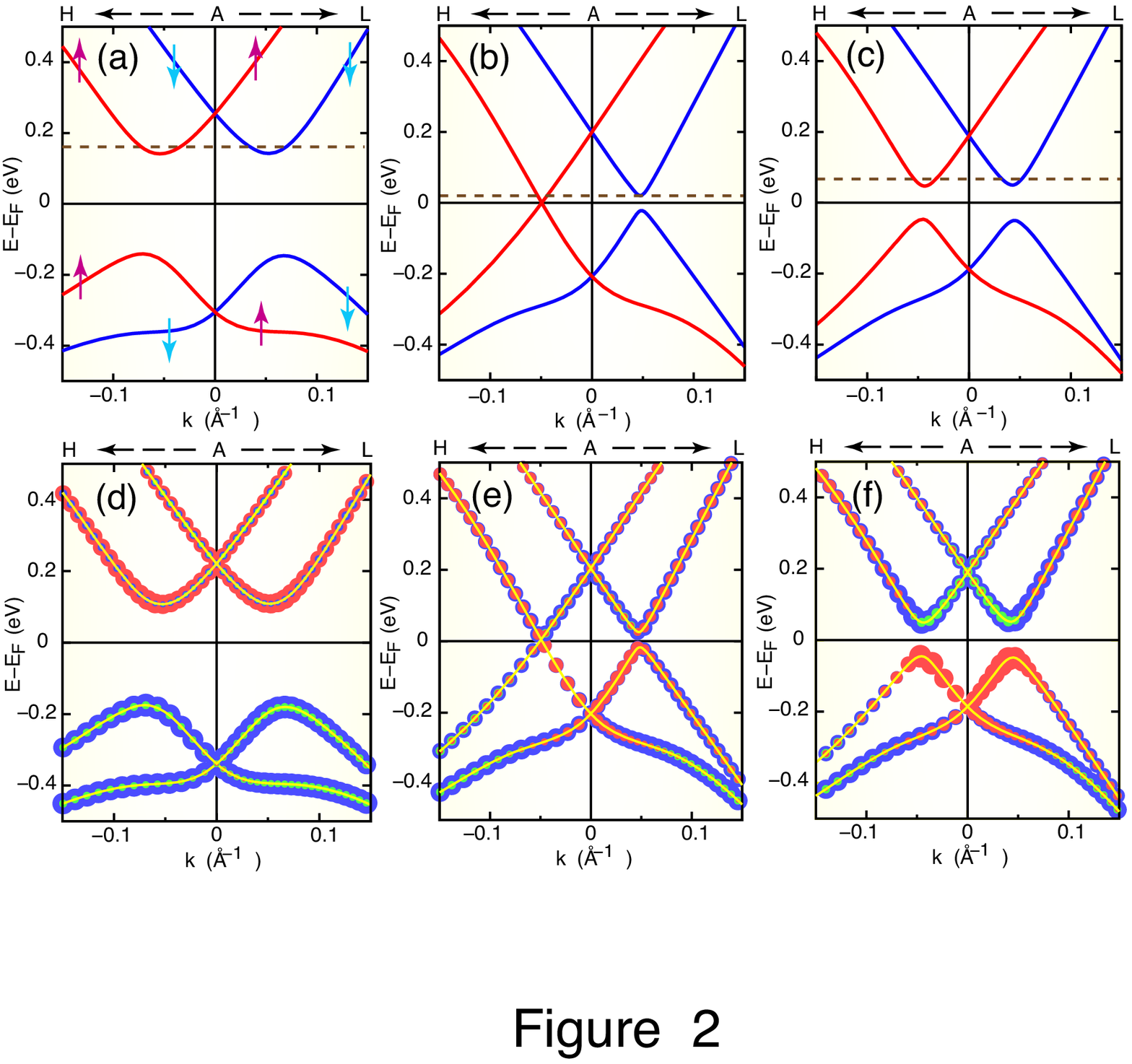}}
\end{center}
\end{figure}

\clearpage

\begin{figure}[pt]
\begin{center}
\rotatebox{0}{\includegraphics[width=7 in]{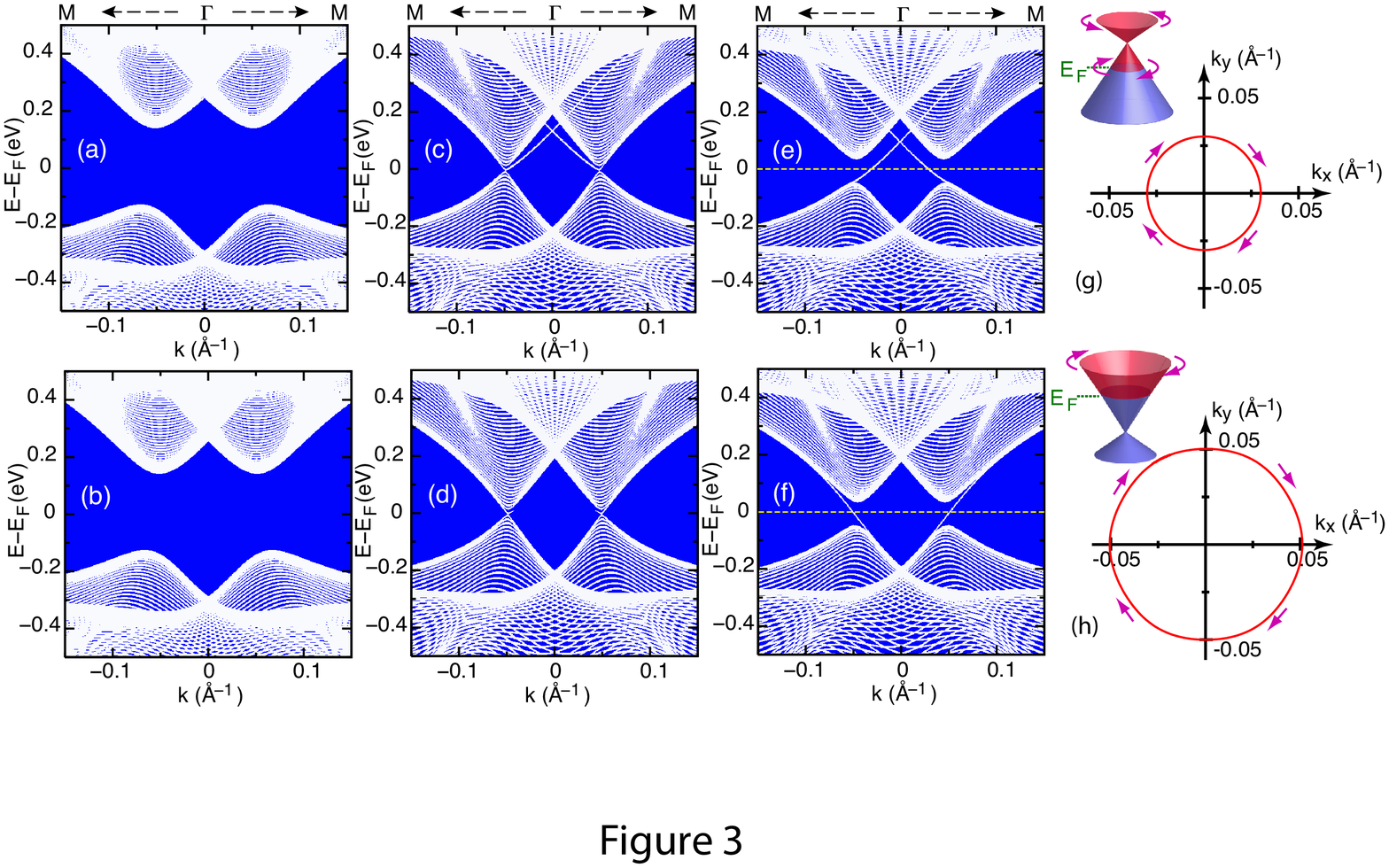}}
\end{center}
\end{figure} 

\clearpage

\begin{figure}[pt]
\begin{center}
\rotatebox{0}{\includegraphics[width=5 in]{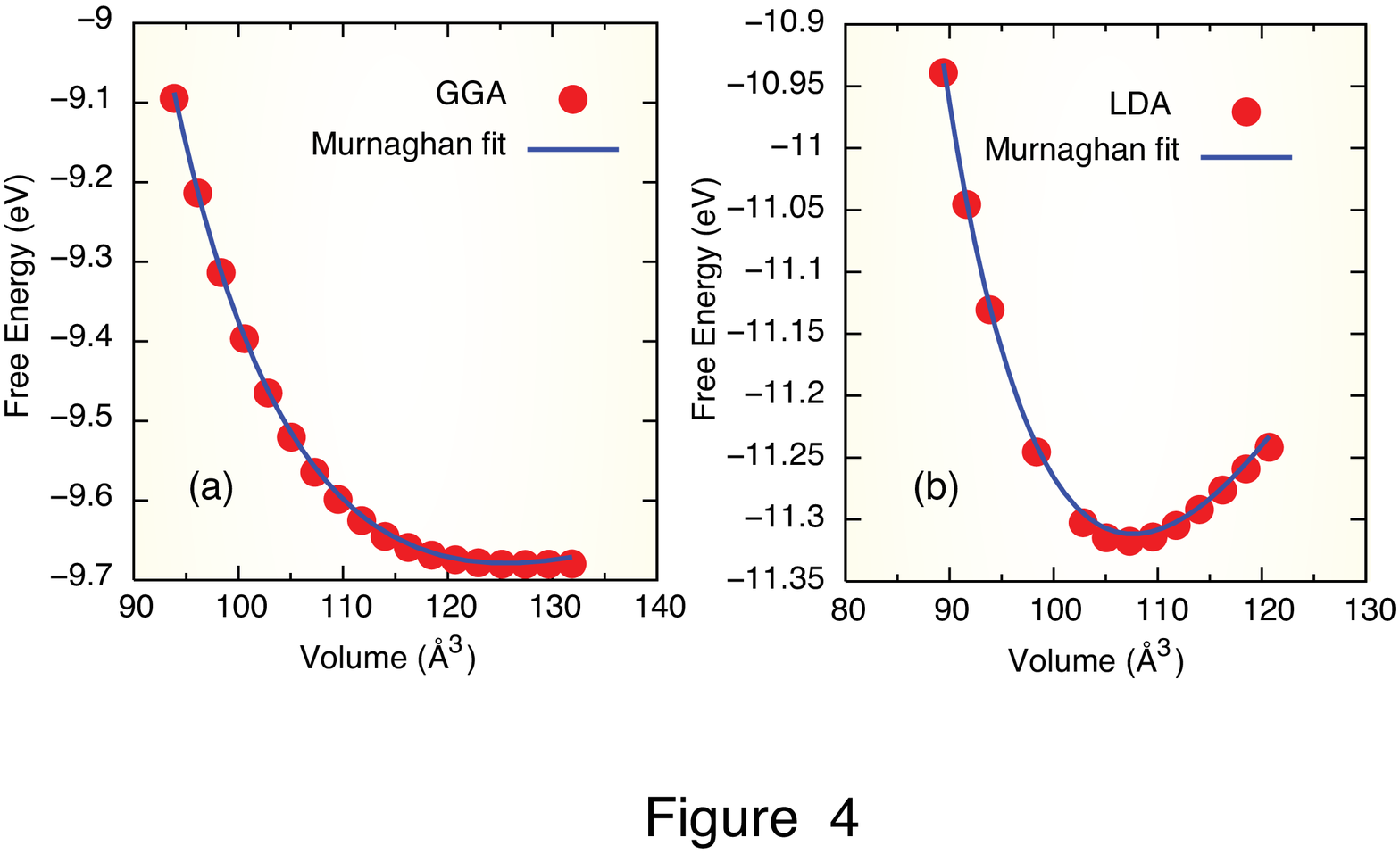}}
\end{center}
\end{figure} 

\newpage

\begin{figure}[pt]
\begin{center}
\rotatebox{0}{\includegraphics[width=5 in]{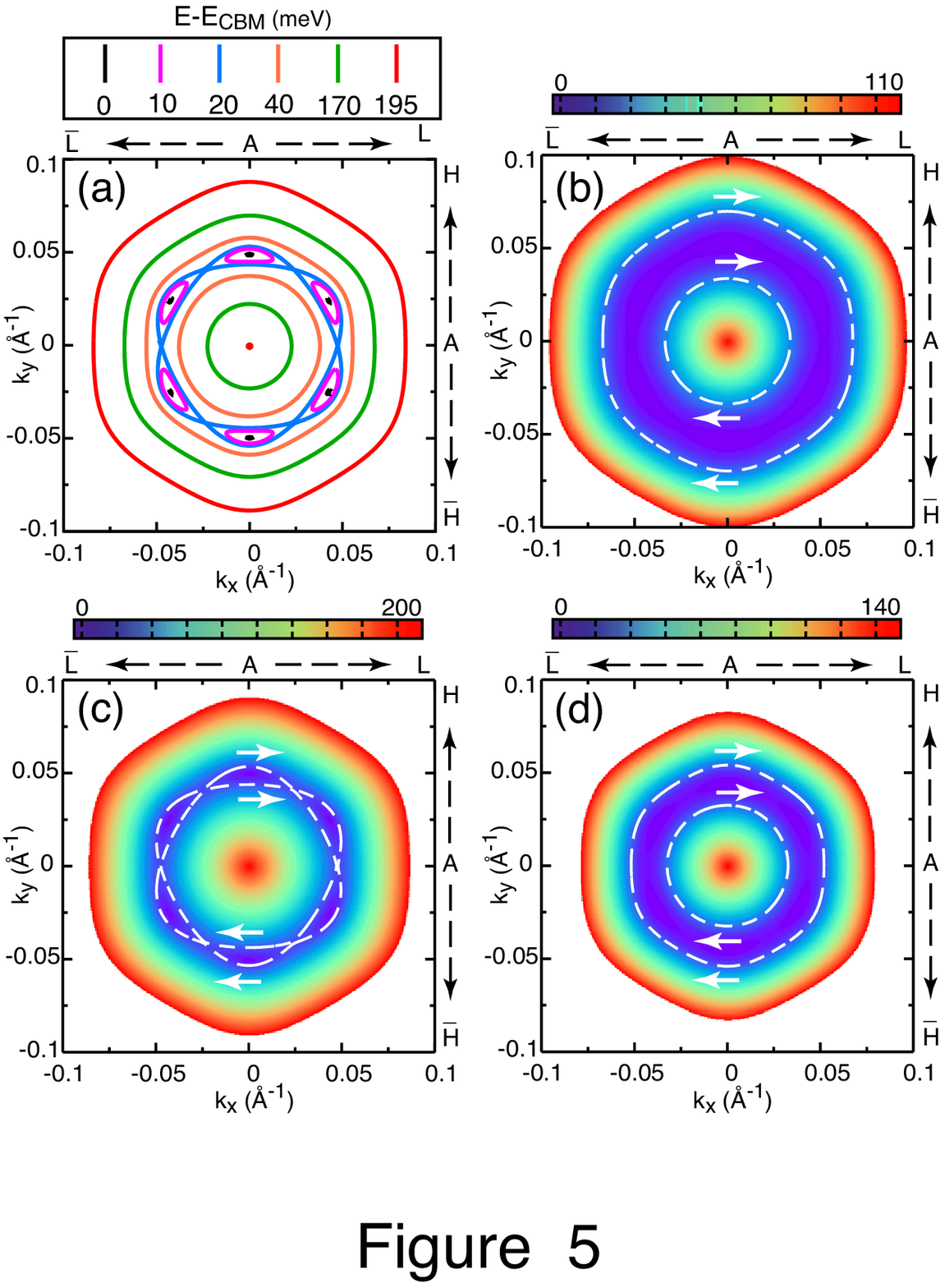}}
\end{center}
\end{figure} 

\end{document}